\documentclass[fleqn,10pt]{wlscirep}
\title{Parametric down-conversion with nonideal and random quasi-phase-matching}

\author[1,+]{Chun-Yao Yang}
\author[1,+]{Chun Lin}
\author[2,+]{Charlotte Liljestrand}
\author[1]{Wei-Min Su}
\author[2]{Carlota Canalias}
\author[1,*]{Chih-Sung Chuu}
\affil[1]{Department of Physics and Frontier Research Center on Fundamental and Applied Sciences of Matters, National Tsing Hua University, Hsinchu 30013, Taiwan}
\affil[2]{Department of Applied Physics, Royal Institute of Technology, Roslagstullsbacken 21, 10691 Stockholm, Sweden}

\affil[*]{cschuu@phys.nthu.edu.tw}

\affil[+]{these authors contributed equally to this work}

\begin{abstract}
Quasi-phase-matching (QPM) has enriched the capacity of parametric down-conversion (PDC) in generating biphotons for many fundamental tests and advanced applications. However, it is not clear how the nonidealities and randomness in the QPM grating of a parametric down-converter may affect the quantum properties of the biphotons. This paper intends to provide insights into the interplay between PDC and nonideal or random QPM structures. Using a periodically poled nonlinear crystal with short periodicity, we conduct experimental and theoretical studies of PDC subject to nonideal duty cycle and random errors in domain lengths. We report the observation of biphotons emerging through noncritical birefringent-phase-matching, which is impossible to occur in PDC with an ideal QPM grating, and a biphoton spectrum determined by the details of nonidealities and randomness. We also observed QPM biphotons with a diminished strength. These features are both confirmed by our theory. Our work provides new perspectives for biphoton engineering with QPM.
\end{abstract}
\begin{document}

\flushbottom
\maketitle

Parametric down-conversion (PDC) is essential to quantum information science in both fundamental studies and potential applications \cite{Aspect82, Briegel98, Knill01}. It allows preparation and manipulation of biphotons with quantum-optical properties of interest \cite{Kwiat95, Ou99, Kuklewicz04, Harris07, Sensarn10, Wagenknecht10, Chuu11}. The key to this flexibility is the technique of quasi-phase-matching (QPM), which employs periodic modulation of the nonlinear coefficient to compensate for the phase mismatch between the interacting waves. The most popular way to implement QPM is by periodically inverting the spontaneous polarization in ferroelectric oxide crystals, so-called periodic poling. In order to obtain a periodic domain-structure an electric field, exceeding the coercive field of the material, is applied to periodic electrodes deposited by lithographic techniques on one of the polar surfaces of the crystals. However, deviations from an ideal structure (for example, nonideal duty cycle or random domain lengths) may be induced during the poling process and lead to complications in parametric interaction \cite{Fejer92, Pelc10, Pelc11, Phillips13}.

In this paper we explore the effects of nonideal duty cycle (nonideal QPM) and random domain lengths (random QPM) in the quantum regime. We experimentally study how they alter the quantum properties of biphotons in PDC. We observed biphotons emerging at the noncritical birefringent-phase-matching (NBPM) wavelength with a spectrum governed by the details of nonidealities and randomness. In addition, we also observed a diminished spectral power density at the QPM wavelength. These unique features are observed in a periodically poled Rb-doped KTP (PPRKTP) crystal with short periodicity and are confirmed by our theory. The nonideal and random QPM structures studied here may also be controlled during the fabrication of the periodically poled crystal. Hence, our work provides not only insights into the interplay between the PDC and nonideal or random QPM structure, but also new perspectives for manipulating biphotons.

\begin{figure} [ht]
\centering
\includegraphics[width=0.5 \linewidth]{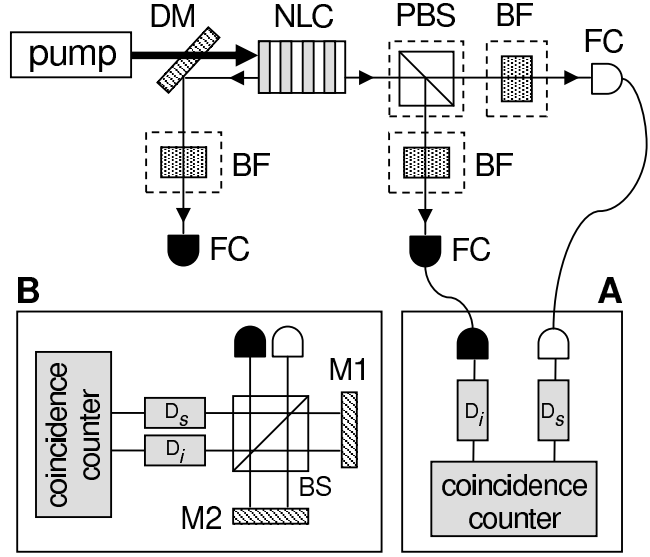}
\caption{\label{fig3_setup} Schematic of experiment setup. NLC: nonlinear crystal, DM: dichroic mirror, PBS: polarizing beamsplitter, BF: bandpass filter, FC: fiber coupler, BS: nonpolarizing beamspliitter, M1/M2: mirrors, D$_s$/D$_i$: single photon detection modules. Accidental coincidence counts from the pump is reduced by long-pass filters (not shown).}
\end{figure}

Our experimental setup is shown in Fig.~\ref{fig3_setup}. The QPM devices employed in this work are fabricated in a RKTP crystal of dimensions $11 \times 6 \times 1 \ {\rm mm}^3$ ($a \times b \times c$ crystallographic axis). Using standard photolithographic technique, the crystal is patterned on the $c^-$ face with three gratings of period $\Lambda$ = 2.112~$\mu$m (grating I), 2.132~$\mu$m (grating II), and 2.152~$\mu$m (grating III). All the gratings have a metal-insulator duty-cycle of $15 \ \%$, a width of 810~$\mu$m, and are separated by a 135-$\mu$m-wide single-domain region. The crystal is periodically poled at room temperature by applying a single 5-ms-long symmetric triangular pulse of 6.8 kV/mm magnitude \cite{Zukauskas13}. 

\begin{table} [ht]
	\centering
	\begin{tabular}{c  c  c  c  c}
		\hline \hline
						& Grating I\ & Grating II\ & Grating III\ & Single domain  \\
		\hline
		$R_b$ & 8.52 & 23.0 & 33.2 & 118  \\
		$R_f$ & 1840 & 4322 & 6599 & 23481  \\
		$R_{b,{\rm QPM}}$ & 0.055 & 0.060 & 0.035 & 0.040  \\
		$R_{f,{\rm QPM}}$ & 16.5 & 25.7 & 13.1 & 21.1  \\
		$R_{b,{\rm NBPM}}$ & 0.16 & 0.40 & 0.56 & 3.13  \\
		$R_{f,{\rm NBPM}}$ & 123 & 426 & 754 & 3608  \\
		$\gamma_1$ & $1.3 \times 10^{-3}$ & $9.4 \times 10^{-4}$ & $7.4 \times 10^{-4}$ & $8.7 \times 10^{-4}$  \\
		$\gamma_2$ & $4.6 \times 10^{-3}$ & $5.3 \times 10^{-3}$ & $5.0 \times 10^{-3}$ & $5.0 \times 10^{-3}$  \\
		$\gamma_3$ & 1.18 & 1.17 & 1.63 & 1  \\
		\hline \hline
	\end{tabular}
	\caption{Pair rates per mW of pump power. The subscripts $b$ and $f$ refer to counter- and co- propagating biphotons, respectively. $R_{b,{\rm QPM}}$ and $R_{f,{\rm QPM}}$ are measured with a bandpass filter of center wavelength 1064 nm and bandwidth 25 nm. $R_{b,{\rm NBPM}}$ and $R_{f,{\rm NBPM}}$ are measured with a bandpass filter of center wavelength 1038 nm and bandwidth 1 nm. The standard deviation is $\pm 10\%$. $\gamma_3$ is normalized to the single-domain value.}
	\label{tab:BiphotonRates}
\end{table}

\begin{figure} [ht]
\centering
\includegraphics[width=0.55 \linewidth]{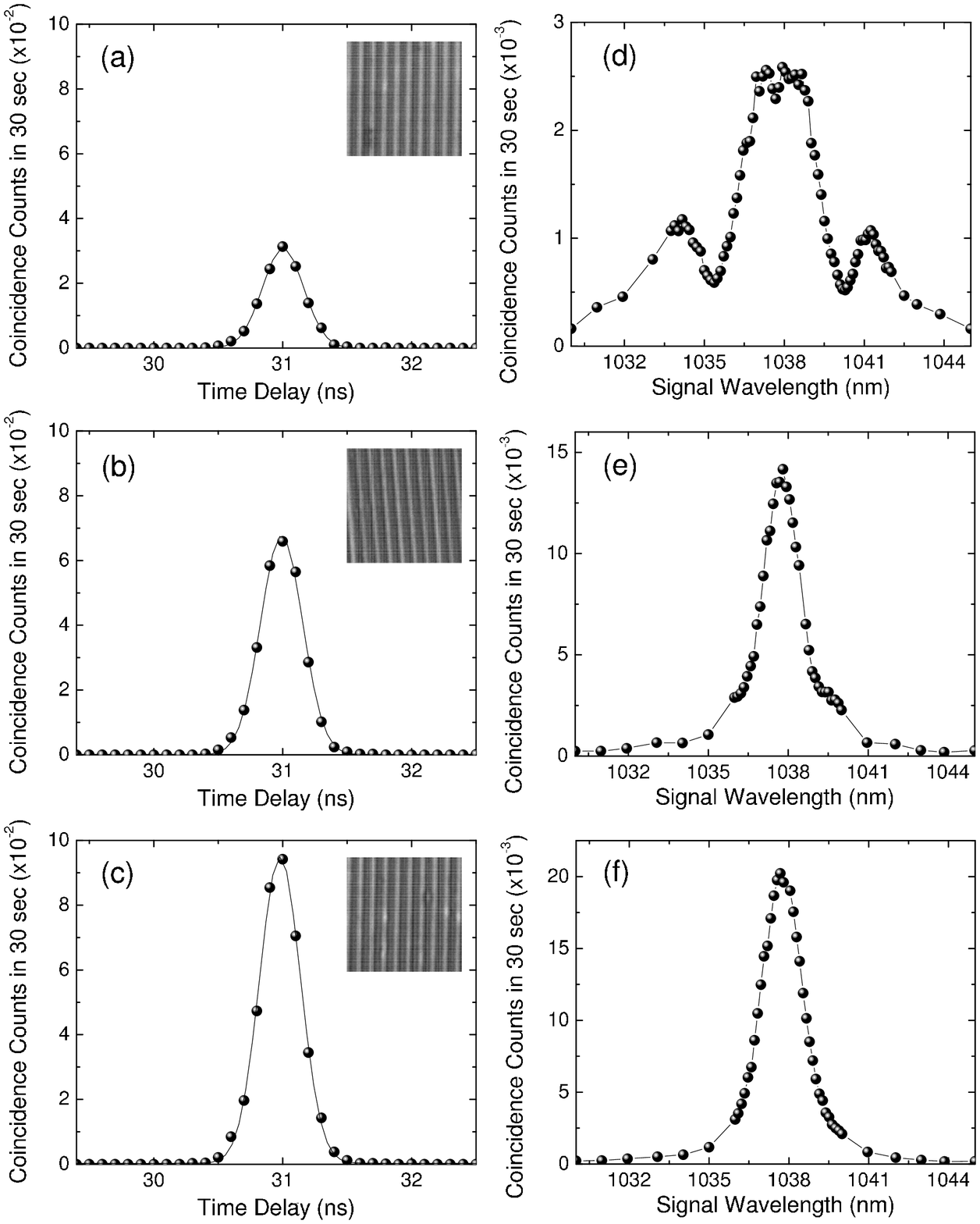}
\caption{\label{fig4_grating_3} (a)-(c): Second-order correlation functions of the biphotons generated from (a) grating I, (b) grating II, and (c) grating III. The curves in (a)-(c) are Gaussian fit to the data points. The insets are images ($20 \times 20 \ \mu$m$^2$) of domain structures on the patterned face. (d)-(f): Spectral power density of the biphotons generated from (d) grating I, (e) grating II, and (f) grating III.}
\end{figure}

We start with the measurement of backward-wave QPM biphotons by pumping the generating crystal with a confocally-focused 532-nm cw laser (Coherent Verdi G5-SLM) near the patterned face (insets in Fig.~\ref{fig4_grating_3}). This would give a QPM signal wavelength of 1074 nm (grating I), 1064 (grating II), 1054 nm (grating III). The counter-propagating signal and idler photons are first passed through bandpass filters (centered at 1064 nm, bandwidth of 25 nm) and guided to single-photon counters (ID Quantique, id400) by multimode fibers (panel A in Fig.~\ref{fig3_setup}). A digital time converter (FAST ComTec, MSC6) then measures the coincidence counts as a function of time delay. By summing the coincidence rates over all time delays, we obtain the pair rate $R_{b,{\rm QPM}}$ per mW of pump power. 

The detected counter-propagating biphotons may also contain ``unexpected'' forward-wave biphotons due to the random and nonideal QPM structures in theperiocially poled crystal. If the forward-wave biphotons have nonzero spectral power density at the QPM wavelentgh, the signal or idler photons can be back-reflected by the crystal end face and contribute to the coincidence counts. The generating crystal is thus tilted at a small angle to reduce the detection of any forward-wave biphotons. To take into account the residual forward-wave biphotons, we measure the pair rates without the bandpass filters $R_b$ and, for the reasons that will be explained later, at the NBPM wavelength $R_{b,{\rm NBPM}}$. The latter is carried out by using a fiber-based bandpass filter centered at 1038 nm (bandwidth 1 nm) in the signal channel and none in idler channel. In addition, we measure the pair rates per mW of pump power of the co-propagating biphotons (in the forward direction) without bandpass filters $R_f$, at the QPM wavelength $R_{f,{\rm QPM}}$ (same filter set as $R_{b,{\rm QPM}}$), and at the NBPM wavelength $R_{f,{\rm NBPM}}$ (same filter set as $R_{b,{\rm NBPM}}$). For these measurements, we use a polarizing beamsplitter to separate the co-propagating signal and idler before counting the photon statistics.  

Tab.~\ref{tab:BiphotonRates} summarizes the measurements of all pair rates in the gratings and the single-domain region. The higher pair rates in the forward direction ($R_f > R_b$) indicates the presence of forward-wave biphotons in addition to the backward-wave QPM biphotons. The ratio $\gamma_1 = R_{b,{\rm NBPM}}/R_{f,{\rm NBPM}}$ is then a measure of the fraction of back-reflected forward-wave biphotons, while $\gamma_2 = R_b/R_f$ also includes the contribution from the backward-wave QPM biphotons. The small $\gamma_1 \approx 0.1~\%$ implies that the detection of back-reflected forward-wave biphotons is suppressed by a factor of $\sim 1000$ due to tilting the crystal. Hence, the nearly equal $\gamma_2$ in the gratings and the single-domain region, of which the latter does not generate backward-wave QPM biphotons, indicates that the pair rate of the backward-wave QPM biphotons is $\sim 1000$ times lower than that of the forward-wave biphotons. This is also evident in the ratio $\gamma_3 = (R_{b,{\rm QPM}}/R_{b,{\rm NBPM}})/(R_{f,{\rm QPM}}/R_{f,{\rm NBPM}})$, which compares the biphoton proportions at the QPM and NBPM wavelengths in the counter-propagating and co-propagating biphotons (normalized to the value in the single-domain region). The small difference between $\gamma_3$ in the grating and single-domain region shows that the generation of backward-wave QPM biphotons is diminished due to the presence of nonideal and random QPM structures.

Because of the short poling periods in our PPRKTP crystal, the observed forward-wave biphotons are not generated by QPM processes. To characterize these biphotons, we first measure their second-order intensity correlations as shown in Fig.~\ref{fig4_grating_3}(a)-(c). Although the correlation functions are broaden by the finite timing resolution (about 0.3 ns) of our detection system, the peaks at the time delay of 31 ns is a clear sign of timely correlated photons. The pair rates per mW of pump power are 108~s$^{-1}$mW$^{-1}$, 242 s$^{-1}$mW$^{-1}$, and 333 s$^{-1}$mW$^{-1}$ at a pump power of 120~$\mu$W for grating I, II, and III, respectively. Corrected for the optical transmittance of 70$\%$ and detector efficiency of $\sim$30$\%$ in each channel, this implies a generation rate of 2,500 s$^{-1}$mW$^{-1}$, 5,575 s$^{-1}$mW$^{-1}$, 7,690~s$^{-1}$mW$^{-1}$ in grating I, II, and III, respectively. 

\begin{figure} [ht]
\centering
\includegraphics[width=0.55 \linewidth]{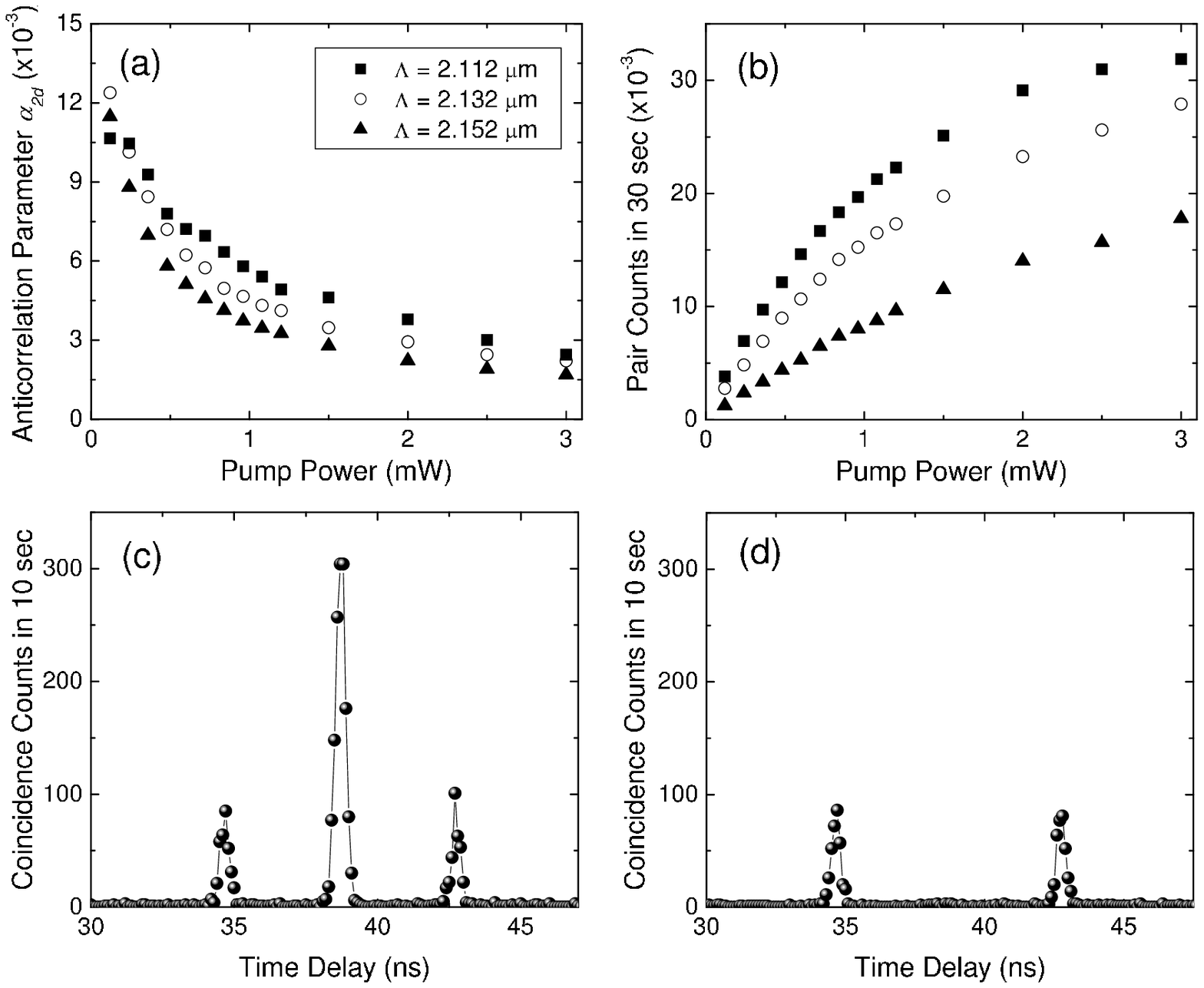}
\caption{\label{fig4_grating_2} (a) Anticorrelation parameter $\alpha_{2d}$ and (b) pair counts as a function of pump power. Single photon counting approaches saturation at higher power. (c) Constructive and (d) destructive interference in coincidence counts between two outputs of the Franson interferometer.}
\end{figure}

We next verify the nonclassical correlation between the signal and idler photons by measuring the anticorrelation parameter $\alpha_{2d} = R_c/(\tau_c R_s R_i)$ \cite{Grangier86}, where $R_c$ is the coincidence rate, $R_s$ and $R_i$ are the singles rate of the signal and idler photons, respectively, and $\tau_c$ is the coincidence window. Fig.~\ref{fig4_grating_2}(a) and (b) show the measured $\alpha_{2d}$ and pair rate as a function of pump power, respectively. For a classical (random) source, $\alpha_{2d}$ is equal to 1. The measured $\alpha_{2d}$ are greater than the classical limit, confirming the nonclassical correlation between the signal and idler photons. 

The time-energy entanglement of the biphotons is also demonstrated by sending the signal and idler photons into a Franson-type interferometer \cite{Franson89} composed of two spatially-separated modes in an unbalanced Michelson interferometer (panel B in Fig.~\ref{fig3_setup}). The biphoton wavefunctions corresponding to signal and idler photons both taking the long arms and both taking the short arms are indistinguishable and lead to interference. Based on the constructive and destructive interference in the coincidence counts between the interferometer outputs (central peaks in Fig.~\ref{fig4_grating_2}(c) and (d), respectively), the fringe visibility is estimated to be $98.0 \pm 0.7 \ \%$, $98.2 \pm 0.5 \ \%$, $97.2 \pm 0.2 \ \%$ for grating I, II, II, respectively, and exceed the classical limit of $50 \ \%$. 

We identify the type of parametric process that generates the forward-wave biphotons by measuring the spectral power density at the signal wavelength. The signal photons are passed through a tunable band-pass filter before entering the single-photon detector. The coincidence counts between the signal and idler photons is then measured for various center wavelengths of the filter at a pump power of 3 mW. As shown in Fig.~\ref{fig4_grating_3}(d)-(f), the spectral power density for all three gratings are centered at the NBPM wavelength (1038 nm). In addition, the polarizations of the signal photons (along the crystal \textit{z}-axis) and idler photons (along the crystal \textit{y}-axis) are also verified by a polarizer to be the same as in the NBPM phase-matching. The biphotons are therefore of the NBPM type. This will not be expected for PDC with ideal QPM condition. Moreover, the variation of spectrum between different gratings and the deviation from a sinc$^2$ function reveal that the nonideal or random QPM structure can alter the properties of the biphotons in different manners. 

To understand the physics behind our observations, we develop our theory in Heisenberg picture. We first consider PDC in the presence of nonideal duty cycle (nonideal QPM), where a periodically poled generating crystal with a duty cycle $D$ deviated from the ideal value of 0.5 is pumped by a single-frequency pump $\omega_p$. The normalized function form of such spatially varying nonlinear coefficient can be described by the Fourier series $\tilde{d}(z) = d(z)/d_{\rm eff} = \sum^{\infty}_{m = - \infty} 2 D {\rm sinc} (\pi m D) {\rm exp} (i K_m z)$ with $m$th-harmonic grating vector $K_m = 2 \pi m / \Lambda$ and $\Lambda$ the poling period. With the down-converted signal and idler fields described by $a_s (\omega, z)$ and $a_i (\omega_i = \omega_p - \omega, z)$, the coupled equations for their slowly varying envelopes $b (\omega, z) = a (\omega, z) {\rm exp} [-i k(\omega) z]$ are
\begin{eqnarray}
\frac{\partial b_s}{\partial z} &=& i \kappa \tilde{d}(z) b^{\dagger}_i \exp[i \Delta k (\omega) z], \nonumber \\
\frac{\partial b_i^{\dagger}}{\partial z} &=& i \epsilon \kappa \tilde{d}^*(z) b_s \exp[- i \Delta k (\omega) z].
\label{eq:coupled_eq}
\end{eqnarray}
Here, $\kappa$ is the coupling constant, $\Delta k (\omega) = k_p (\omega_p) - k_s (\omega) - \epsilon k_i (\omega_i)$ is the phase-mismatch, and $\epsilon = +1$ (forward-wave interaction) or $-1$ (backward-wave interaction). For small parametric gain, we obtain the output fields $a_s (\omega, L)$ and $a_i (\omega_i, z_{\rm out})$,
\begin{eqnarray}
a_s (\omega, L) &=& A(\omega) a_s (\omega, 0) + B(\omega) a^{\dagger}_i (\omega_i, L-z_{\rm out}), \nonumber \\
a_i^{\dagger} (\omega_i, z_{\rm out}) &=& C(\omega) a_s (\omega, 0) + D(\omega) a_i^{\dagger} (\omega_i, L-z_{\rm out}), 
\label{eq:coupled_eq_sol1}
\end{eqnarray}
where $z_{\rm out} = L (\epsilon+1)/2$, $A(\omega) = {\rm exp} [i k_s(\omega) L]$, $C(\omega) = B^*(\omega) {\rm exp} [ i (k_s(\omega)+(\epsilon-1) k_i(\omega_i)) L ]$, $D(\omega) = {\rm exp} [- \epsilon i k_i(\omega_i) L]$, and
\begin{eqnarray}
B(\omega) &=& \left( \sum_{m \neq 0} B_m (\omega) + B_0 (\omega) \right) \exp \left\{i \left[ \frac{\Delta k(\omega)}{2} + k_s(\omega) + k_i(\omega_i) \right] L \right\}, \label{eq:B} \\
B_m (\omega) &=& 2 i \kappa L \ \frac{{\rm sin} (\pi m D)}{\pi m} \ {\rm sinc} \left[ \frac{(\Delta k (\omega) + K_m)L}{2} \right], \nonumber \\
B_0 (\omega) &=& 2 i \kappa L D \ {\rm sinc} \left[ \frac{\Delta k (\omega) L}{2} \right] \nonumber.
\end{eqnarray}

\begin{figure}
\centering
\includegraphics[width= 0.6 \linewidth]{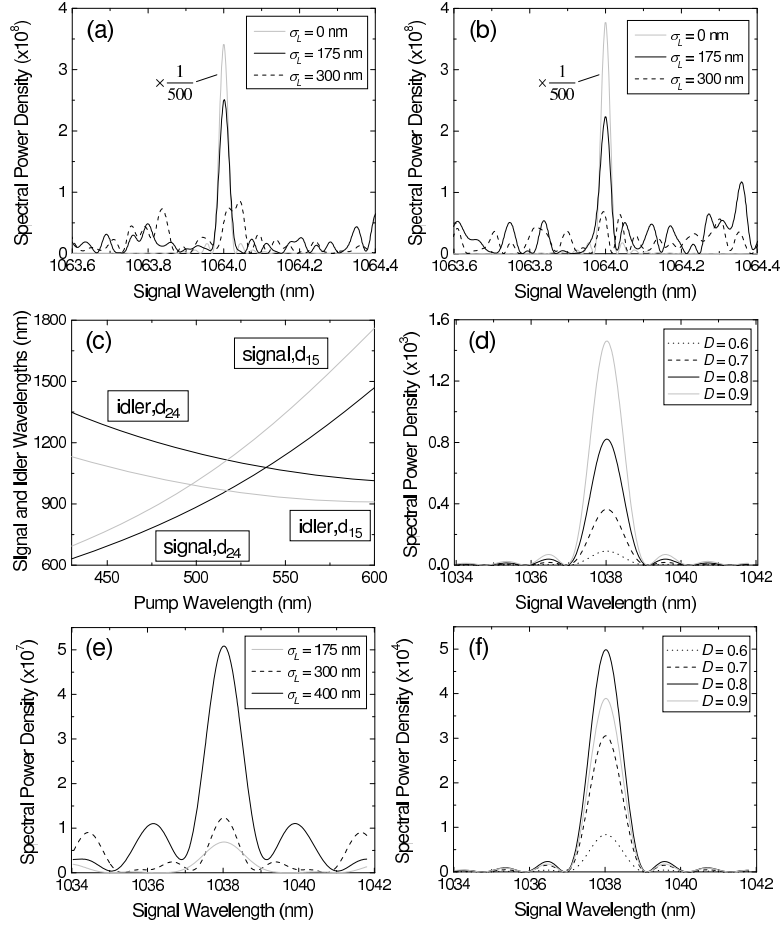}
\caption{\label{fig1_NBPM} (a)-(b): Calculated spectral power density of QPM biphotons for (a) $D = 0.8$ with $\sigma_L =$ 0, 175, 300 nm and (b) $D = 0.5$ with $\sigma_L = $ 0, 175, 300 nm. (c) Wavelengths of NBPM biphotons in KTP crystal as a function of pump wavelength. (d)-(f): Calculated spectral power density of NBPM biphotons for (d) $D = 0.6, 0.7, 0.8, 0.9$ without random phase error, (e) $D = 0.5$ with $\sigma_L =$ 175, 300, 400 nm, and (f) $D = 0.6, 0.7, 0.8, 0.9$ with $\sigma_L = 300$ nm}
\end{figure}

The spectral power density of the biphotons $S(\omega) = |B(\omega)|^2 /2 \pi$ is governed by Eq.~(\ref{eq:B}). If the $n$th-harmonic grating vector $K_n$ compensates the phase-mismatch $\Delta k (\omega_0)$ at frequency $\omega_0$, $B_n (\omega)$ dominates the summation in Eq.~(\ref{eq:B}) and results in $S(\omega) \propto {\rm sinc}^2 [(\Delta k (\omega)+K_m) L/2]$ in the vicinity of $\omega_0$. This leads to biphoton generation at the QPM frequency $\omega_0$ with a strength reduced by a factor of $4 D^2 {\rm sinc}^2 (\pi m D)$ as compared to the ideal case ($D=0.5$). The gray curves in Fig.~\ref{fig1_NBPM} (a) and (b) are the $S(\omega)$ with nonideal ($D=0.8$) and ideal ($D=0.5$) duty cycle in the vicinity of QPM wavelength. These calculations assume a PPKTP crystal with $\Lambda = 2.132 \ \mu$m and a 532-nm pump, which generates backward-wave QPM biphotons at 1064 nm. 

The $B_0(\omega)$ term in Eq.~(\ref{eq:B}), which is independent of $K_m$, is absent in a perfectly poled crystal with ideal duty cycle. However, with the presence of nonideal duty cycle ($D \neq 0.5$), $B_0(\omega)$ is nonzero and has a peak value at the NBPM frequency $\omega_1$ satisfying $\Delta k (\omega_1) = 0$. The conditions for NBPM in a KTP crystal are shown in Fig.~\ref{fig1_NBPM}(c) for pump wavelengths ranging from 420 to 600 nm. These parametric interactions are of type-II and forward-wave type. Their existence allows NBPM biphotons to be generated in the presence of QPM grating. Compared to the QPM biphotons, the $S(\omega)$ of the NBPM biphotons also has a sinc$^2$ shape (this is not true anymore if there is random phase error) but with a magnitude increasing linearly with the duty cycle (Fig.~\ref{fig1_NBPM}(d)). The ratio of the peak spectral power density of the NBPM biphotons to that of the QPM biphotons is $S(\omega_1)/S(\omega_0) = \kappa_1^2/(\kappa_0^2 \ {\rm sinc}^2 (\pi n D))$.

We now include the effect of random domain lengths (random QPM) in PDC by allowing position errors to occur at the domain boundaries where $\tilde{d}(z)$ changes its sign. We describe the left and right boundaries of the $m$th domain by $z_m^l = z_{m,0}^l + \delta z_m^l$ and $z_m^r = z_{m,0}^r + \delta z_m^r$, respectively, where $z_{m,0}^{l,r}$ are the ideal positions of domain boundaries and $\delta z_m^{l,r}$ are position errors which have a normal distribution with a standard deviation of $\sigma_L$. With the domain boundaries in Eq.~(\ref{eq:coupled_eq}) defined by $z_m^{l,r}$, we can calculate $S(\omega)$ numerically for various $\sigma_L$ and duty cycles.

The solid (black) and dashed curves in Fig.~\ref{fig1_NBPM}(b) show the spectral power density of the QPM biphotons with random domain lengths and an ideal duty cycle. The $S(\omega)$ manifests a magnitude and shape that depend on the exact distribution of random phase errors. In general, as the standard deviation increases, the reduction at the QPM wavelength, the enhancement at other wavelengths, and the deformation in shape become more pronouncedly. The emergence of NBPM biphotons is again revealed in $S(\omega)$ (Fig.~\ref{fig1_NBPM}(e)) with its magnitude at the NBPM wavelength increasing with the standard deviation. Unlike the PDC with only nonideal QPM, the frequency and magnitude of the secondary peaks in $S(\omega)$ can deviate significantly from an ideal sinc$^2$ function. This is different as compared to the PDC with random domain lengths and nonideal duty cycle, in which $S(\omega)$ of the NBPM biphotons exhibits a shape resembling the sinc$^2$ function (Fig.~\ref{fig1_NBPM}(f)). In contrast, the $S(\omega)$ of the QPM biphotons (Fig.~\ref{fig1_NBPM}(a)) has a similar characteristic as in the PDC with only random domain lengths (Fig.~\ref{fig1_NBPM}(b)). 

\begin{figure}
\centering
\includegraphics[width= 0.6 \linewidth]{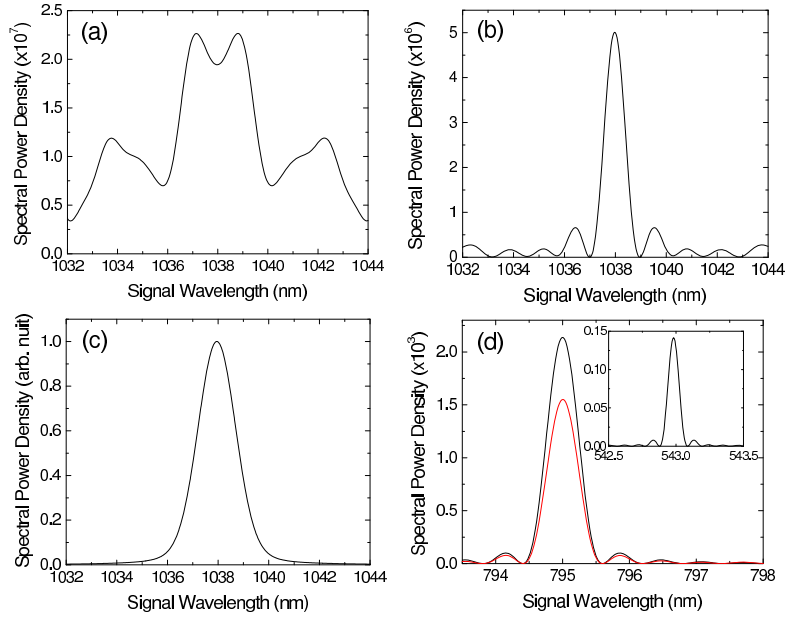}
\caption{\label{fig5_simulation} (color online) Simulated spectral power density of (a) NBPM biphotons in backward-wave PDC with $\Lambda = 2.112 \ \mu$m, $D = 0.5$, $\sigma_L =$ 450 nm, (b) NBPM biphotons in backward-wave PDC with $\Lambda = 2.132 \ \mu$m, $D = 0.5$, $\sigma_L =$ 600 nm, (c) NBPM biphotons in backward-wave PDC with an ideal sinc$^2$ function broaden by finite spectral resolution of 1 nm, and (d) QPM biphotons in forward-wave PDC with $\Lambda = 8.9 \ \mu$m, $D = 0.6$, $\sigma_L =$ 900 nm (red) and without poling errors (black). The inset shows the spectral power density of the NBPM biphotons.}
\end{figure}

For the emergent NBPM biphotons, the presence of random domain lengths thus deforms or broadens their spectrum while the nonideal duty cycle preserves the sinc$^2$ shape of their spectrum. Based on these observations, we try to explain the origins of our spectral measurements in Fig.~\ref{fig4_grating_3} as follows. The biphoton spectrum with grating I (Fig.~\ref{fig4_grating_3}(d)) reveals a broaden width and an enhanced strength at non-NBPM wavelengths. This is a clear sign that the random domain lengths is the dominant effect. To verify this, we carry out simulations of biphoton spectrum assuming an ideal duty cycle and random domain lengths with a standard deviation of 450 nm. As shown by the example in Fig.~\ref{fig5_simulation}(a), the simulated spectrum catches most features of the observed spectrum in Fig.~\ref{fig4_grating_3}(d). The biphoton spectrum with grating II (Fig.~\ref{fig4_grating_3}(e)) has a less deformed shape and seems to be dominant by nonideal duty cycle. However, the enhanced magnitude of the side peaks (blurred by the finite resolution) reveals again the significance of random domain lengths. This is supported by the simulation in Fig.~\ref{fig5_simulation}(b), where we assume an ideal duty cycle and random domain lengths with a standard deviation of 600 nm. Enhanced side peaks can be seen next to the center peak. The biphoton spectrum with grating III (Fig.~\ref{fig4_grating_3}(f)), in contrast to the other spectrum, has a nearly non-deformed shape and thus implies the significant role of nonideal duty cycle. The deviation from an ideal sinc$^2$ function is due to the finite resolution of our spectral measurement. By taking into account the bandwidth of our bandpass filter and assuming only nonideal duty cycle, the simulation in Fig.~\ref{fig5_simulation}(c) reproduces the shape and width of the measured spectrum in Fig.~\ref{fig4_grating_3}(f).

In summary, this work reports a first study of PDC subject to nonideal and random periodically-poled structures in the generating crystal. The observations of QPM biphotons with a reduced strength and NBPM biphotons with a spectrum dependent on the nonidealities and randomness are both confirmed by our theory, which considers PDC with nonideal duty cycle and random domain lengths. Our study suggests that high quality of periodic poling is necessary to realize backward-wave PDC with high generation efficiency and to avoid ``unwanted'' parametric processes in PPRKTP. Although this work focuses on backward-wave PDC, the theory presented here can also be applied to study other PDC processes with nonideal duty cycle and random domain lengths. For example, we consider the forward-wave PDC of first-order and type-II QPM. Such parametric process has been used to generate polarization-entangled photons \cite{Kuklewicz04}. In Fig.~\ref{fig5_simulation}~(d), we show the simulated spectrum of QPM biphotons (red curve) subject to realistic poling errors \cite{Pelc11}. The peak spectral power density is reduced by about 20~\% compared to the ideal case (black curve). The poling errors also lead to the emergence of NBPM biphotons (inset) with a peak spectral power density about 10~\% of that of QPM biphotons. To the extent that the duty cycle and the standard deviation of domain lengths can be controlled, one may be able to manipulate the biphoton spectrum via nonideal or random QPM. Compared to an ideal structure, random periodically-poled structures can allow efficient frequency conversion over a wide bandwidth without the need of stringent phase-matching condition \cite{Baudrier-Raybaut04, Fischer06, Bahabad08}. Our work thus provides insights into the interplay between the PDC and nonideal or random QPM, as well as new perspectives for manipulating biphotons.

\section*{Acknowledgments}

This work was supported by the Taiwan Ministry of Science and Technology. We acknowledge experimental assistance from C.-C. Lin, C.-Y. Wang, and C.-Y. Cheng.

\section*{Author contributions statement}

C.Y.Y, C.L, and C.S.C conceived the experiment, C.L. and C.C. fabricated the QPM devices, C.Y.Y, C.L, and W.M.S. conducted the experiment, C.Y.Y, C.L, W.M.S., and C.S.C analyzed the results. All authors reviewed the manuscript. 

\section*{Additional information}

\noindent \textbf{Competing financial interests} The authors declare that they have no competing financial interests.


\end{document}